\documentclass[12pt]{iopart}

\usepackage{iopams}  
\usepackage{graphicx}

\begin{document}

\title{Investigating the radial flow like effects using identified triggered correlation in pp collisions at $\sqrt{s}$ = 7 TeV}

\author{Debojit Sarkar$^1$$^,$$^2$, Supriya Das$^2$ and Subhasis Chattopadhyay$^3$}
\address{$^1$ {Laboratori Nazionali di Frascati, INFN, Frascati, Italy}}
\address{$^2$ Bose Institute, CAPSS, Block EN, Sector 5, Kolkata 700091, India.}
\address{$^3$ Variable Energy Cyclotron Centre, HBNI, 1/AF-Bidhannagar, Kolkata-700064, India}
\ead{debojit03564@gmail.com (Debojit Sarkar)}

\begin{abstract}
An inclusive baryon to meson enhancement with increase in multiplicity has been observed in pp collisions at $\sqrt{s}$= 7 TeV.  Such a striking feature
of the data can be explained by approaches based on hydrodynamics and  multi-parton interactions (MPI) coupled with color reconnection (CR) mechanism. In this paper, we investigate the multiplicity evolution of the charged particle yields associated with pions and protons selected from the intermediate $p_{T}$ region where the inclusive baryon to meson enhancement has been observed. The study has been peformed using EPOS 3 (hydrodynamics) and PYTHIA 8 (MPI with CR) event generators in pp collisions at 7 TeV. We find that the response of the individual pion and proton triggered correlation towards these two mechanisms is different and can be used to disentangle the effect of one from the other.  The current
study can, therefore,  provide important insights on the origin of radial flow like effects in
high multiplicity pp collisions at the LHC energies.

\end{abstract}

\submitto{\jpg}
\maketitle

%
\vspace{2pc}

\section{Introduction}

Observation of radial flow like effects on spectra \cite {p_pi_enhancement_pPb,alice_pPb_radialflow,alice_protontopion_SQM2016,CMS_strangespectra_rf_paper}, ridge like structures \cite {CMS_pp_Ridge} \cite {alice_pPb_double_ridge}, mass ordering of ($v_{2}$) \cite {pPb_mass_ordering} \cite {CMS_pp_v2}, strangeness enhancement \cite {alice_strangenessenhance_pp7}, \cite {alice_strangenessenhance_pPb}  etc in high multiplicity pp and p-Pb collisions at the the LHC energies have triggered the debate on the applicability of hydrodynamics in small collision systems. All these observations have striking similarity with similar measurements in heavy ion collisions \cite {HI_review,PbPb_spectra,Au_Au_p_to_Pi,Pb_Pb_P_to_Pi} where these features have been attributed to hydrodynamical response of a strongly interacting system to the initial spatial anisotropy. However, for small collision systems, the idea of thermalization is debatable \cite {intro_12} and the absence of jet quenching in p-Pb collisions at the LHC energy \cite {RpPb} questions the formation of deconfined medium in small systems. Nevertheless, assuming the applicability of hydrodynamics
\cite {Shuryak_radialflow,flow_pp_pghosh,flow_pp_victor,pp_glauber_enterria} one can explain several experimental observations in high multiplicity classes of small collision systems at the LHC energies \cite{bozek_pPb,epos_ridgein_pp,epos_massordering_flow_pPb,epos_radialflow_spectra_pPb}.
For example, simulations
based on EPOS 3 which includes event by event 3+1 D
hydrodynamic evolution can provide a good explanation of ridge \cite {epos_ridgein_pp,epos_massordering_flow_pPb},  mass ordering of the elliptic flow co-efficients ($v_{2}$) of identified particles \cite {epos_massordering_flow_pPb}, hardening of spectra with multiplicity and the trend of baryon to meson enhancement at the intermediate transverse momentum ($p_{T}$) in p+p and p+Pb collisions at the LHC energies \cite {p_pi_enhancement_pPb,alice_pPb_radialflow,alice_protontopion_SQM2016, epos_radialflow_spectra_pPb,
pp_pPb_meanpt_ALICE}.

But, alternative approaches based on multi parton intercations (MPI) with color reconnection \cite {ortiz_radialflow,ortiz_PYT_EPOS}, incohrent parton scattering with coalescence model of hadronization \cite {AMPT_pp_pPb_ridge,AMPT_pPb_flow,flow_AMPT_dsarkar}, correlated emission from glasma flux tubes (CGC approach) \cite{CGC_1,CGC_2,PYTHIA_massordering_Tribedy} and geometrical (eikonal) estimation of the eccentricity of the partonic overlap region \cite{pp_glauber_enterria} can also reasonably explain the collective signatures in high multiplicity classes of small collision systems. By far, no experimental observable has been able to convincingly discriminate the different scenarios. Therefore, the debate, on whether the collective behaviors observed in high multiplicity pp collisions are due to  hydrodynamic response to intial spatial anisotropy or some alternative approach mimicking the final state collectivity is still on. In this paper, we investigate the multiplicity evolution of pion and proton triggered correlation at intermediate $p_{T}$ using EPOS 3 and PYTHIA 8 event generators to investigate the origin of baryon to meson enhancement in high multiplicity pp collisions at the LHC energy \cite {epos_radialflow_spectra_pPb,ortiz_radialflow,ortiz_PYT_EPOS}.

In one of our recent publications \cite {trigdilution_EPOS_pPb}, the effect of hydrodynamical flow on the pion and proton triggered correlation at intermediate $p_{T}$ has been studied
using the EPOS 3 event generator for p-Pb collisions at $\sqrt{s_{NN} }$ = 5.02 TeV. The pion and proton triggers were selected from intermediate  $p_{T}$ region (2.0  $<p{_T}< $4.0 GeV/$\it{c}$) where the inclusive proton to pion enhancement has been observed and can be explained in terms of radial flow in the context of EPOS 3 \cite {trigdilution_EPOS_pPb}. As baryon production is favored in hydroynamic framework, a stronger suppression in the baryon triggered jetlike correlation with multiplicity has been observed  compared to that of mesons - an effect commonly referred to as ”trigger dilution” \cite{trigdilution_EPOS_pPb,phenix_triggerdilution,star_triggerdilution}.

Now, PYTHIA 8 with color reconnection can also qualitatively explain the proton to pion enhancement at intermediate $p_{T}$ \cite {ortiz_radialflow} as shown in Fig 1.  In addition, it can reasonably explain the experimentally observed mass and multiplicity dependence of mean $p_{T}$, hardening of spectra, narrowing of balance function with multiplicity etc  in pp collisions at 7 TeV \cite {epos_radialflow_spectra_pPb,ortiz_radialflow,BF_ALICE_pp}. By construction, the hard and soft components of the interaction are strongly coupled in the color reconnection mechanism and it plays a major role in generating the radial flow like signatures in high multiplicity pp collisions \cite {ortiz_PYT_EPOS}. For example, the blast wave model \cite{blast_wave} can describe the spectra in high multiplicity pp events only in the presence of CR mechanism and the fitting gets better with increase in $p_{T}^{jet}$ \cite {ortiz_PYT_EPOS}. This indicates that in  PYTHIA 8, the jet induced radial flow mimicks   the experimentally observed collective like behaviors in high multiplicity pp collisions at the LHC energy.

It is important to note that the construction of the jet part is similar in both EPOS 3 and PYTHIA 8 event generators. The additional underlying mechanisms such as hydrodynamical flow in EPOS 3 and color reconnection in PYTHIA 8 are responsible for generating the collective like effects in high multiplicity pp collsions. So, any possible difference in the multiplicity evolution of the pion and proton triggered jetlike yields in these two models can be used to disentangle the effect of hydrodynamics from that of color erconnection and may shed light on the particle production mechanisms at intermediate $p_{T}$  region of small collision systems.

 The paper is organized as follows. In the next section we give a brief description of EPOS 3.107 and PYTHIA 8 event generators. The method of extraction of the jetlike yields from background subtracted correlation function is discussed in section 3. The discussions on results are performed in section 4.

\section{EPOS 3  and PYTHIA 8 event generators}
{\it The EPOS 3 model} -
EPOS 3 basically contains a 3+1 D hydrodynamical approach based on flux tube initial conditions \cite {epos_radialflow_spectra_pPb,epos_model_descrip,parton_gribov_Regge_Th}. 
The formalism used in this model is referred as "Parton based Gribov Regge Theory" and explained 
in detail in \cite {epos_massordering_flow_pPb,parton_gribov_Regge_Th}.  After multiple scatterings the final state partonic system consists of mainly longitudinal colour flux tubes carrying transverse momentum of the hard scattered partons in transverse direction. Depending on the energy of the string segments and local string density - the high density areas form the "core" and the low density areas form the "corona" \cite {epos_core_corona_sep}. The strings in the core thermalize and then undergo hydrodynamical expansion and finally hadronize to constitute the soft part of the system. The strings in the corona hadronize by Schwinger's mechanism and constitute the jet part of the system. As EPOS 3 can describe the experimentally measured inclusive $p_{T}$ distribution of jets in pp collisions at 7 TeV \cite{epos_radialflow_spectra_pPb} as well as collective like behaviors in high multiplicity pp and p-Pb collisions at the LHC energies \cite {alice_pPb_double_ridge,epos_ridgein_pp,epos_massordering_flow_pPb,epos_radialflow_spectra_pPb}, it would be interesting to investigate the effect of the hydrodynamical flow on the near side jet like yields at the intermediate $p_{T}$ region where the inclusive proton to pion enhancement has been observed.

\begin{figure}
\begin{center}
\includegraphics[height=6.0 cm, width=9.0 cm]{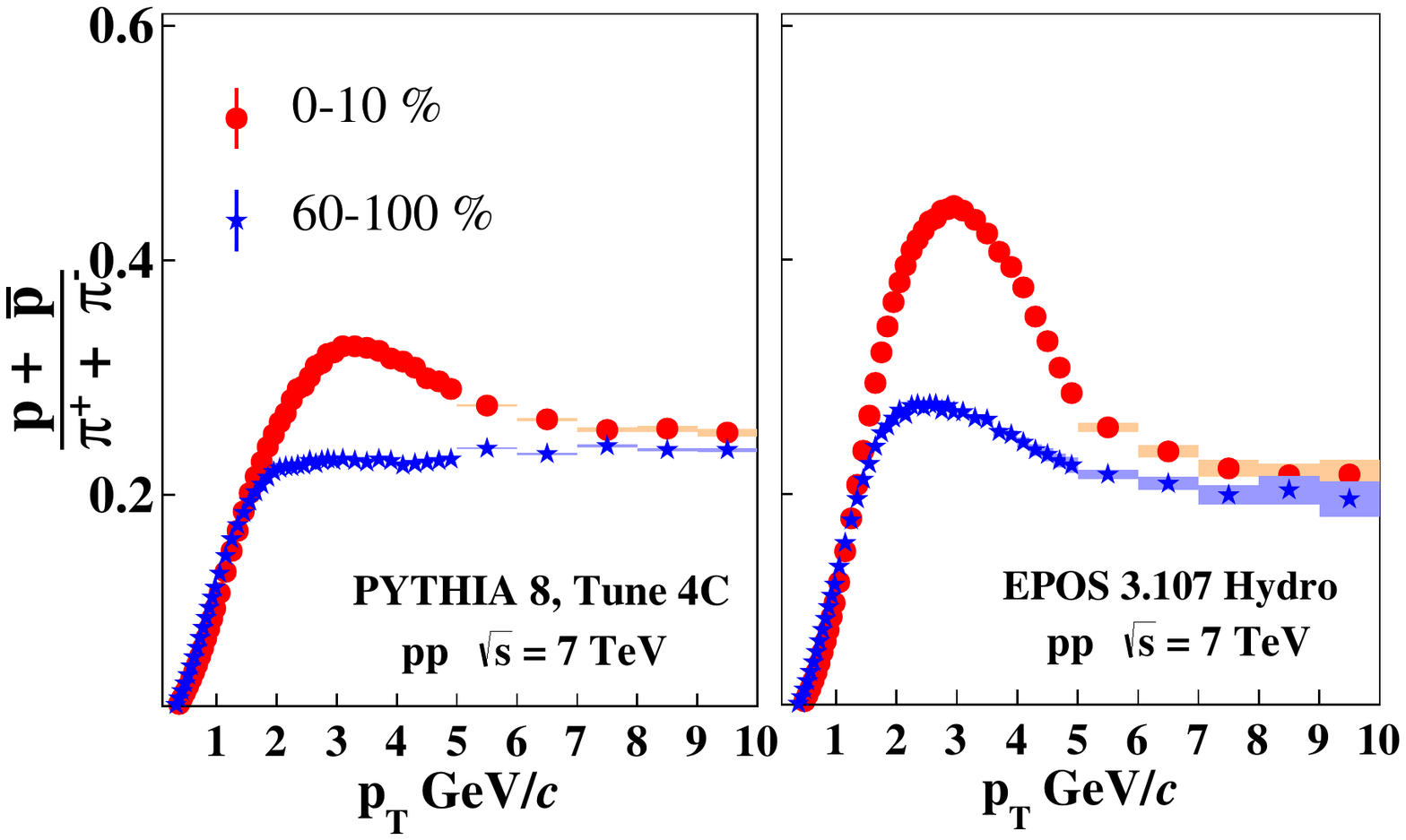}
\caption{[Color online] Inclusive proton over pion ratio as obtained from PYTHIA 8 (left) and EPOS 3 (right)
in 0-10\% and 60-100\% event classes of pp collisions at $\sqrt{s}$ = 7 TeV.}
\label{inclusive-ratio}
\end{center}
\end{figure}

\vspace{0.4cm}
{\it The PYTHIA 8 model} -
PYTHIA is a multi parton interaction \cite {MPI_PYTHIA} based event generator for pp collisions. In PYTHIA, each collision is described by leading-order PQCD calculations, intial and final state parton radiation, multiple parton-parton interactions (MPI), beam remnants and Lund string fragmentation model of hadronization \cite {ortiz_PYT_EPOS}. The MPI can reasonably explain the multplicity distribution and correlation between transverse sphericity and multiplicity in pp collisions at LHC energy. But to explain the collective like behaviors as observed in high multiplicity pp collisions at LHC energy, the color reconnection mechanism \cite {CR_PYTHIA} is required. In this mechanism, final partons from independent hard scatterings are color connected to generate a transverse boost \cite {ortiz_radialflow}. This effect is more prominent for events with several partonic scatterings as it generates a large transverse boost mimicking the radial flow like effect in hydrodynamics. Though the origin of this boost like effect in CR mechanism is completely different compared to the hydrodynamics, this can explain the collective like behaviors such as hardening of spectra with multiplicity, inclusive baryon to meson enhancement at intermediate $p{_T}$, mean $p{_T}$ vs multiplicity and it's mass dependence, narrowing of balance function with multiplicity \cite{BF_ALICE_pp} as observed in the pp collisions at LHC energy and can be explained within the hydrodynamic approach  \cite {epos_radialflow_spectra_pPb} \cite {ortiz_radialflow}. In this paper, we demonstrate that the multiplicity evolution of the identified triggered correlation at the intermediate $p_{T}$ can disentangle the color reconnection from hydrodynamic flow in small collision systems and provide insight into the particle production mechanisms at intermediate $p_{T}$ region.

\section{Analysis Method}
Two particle correlation technique has been extensively used in high-energy physics for extracting the properties of the system and has been discussed in detail in \cite {trigdilution_EPOS_pPb}. The $p_{T}$ range of trigger and 
associated particles are 2.0 $<p{_T}<$ 4.0 GeV/$\it{c}$ and 0.5 $<p{_T}<$ 4.0 GeV/$\it{c}$ respectively and the correlation function has been constructed with a $p_{T}$ ordering ($p_{T}^{assoc} < p_{T}^{trigger}$). The pseudo-rapidity of the particles are 
restricted within -0.8$<\eta<$0.8. 

Both p/$\pi$ ratio and correlation analysis have been performed by dividing the entire minimum bias events
into four multiplicity classes based on the total amount of charged particles producd within 2.8  $<\eta<$5.1 or −3.7 $<\eta<$ −1.7. This corresponds to the acceptance range of ALICE VZERO-A and VZERO-C detector and used for multiplicity class determination by the ALICE collaboration \cite {pp_pPb_meanpt_ALICE}. The multiplicity classes are denoted as 60-100$\%$, 40-60$\%$, 10-40$\%$, 0-10$\%$ from the lowest to the highest multiplicity.
Two particle correlation functions with proton triggers as obtained from PYTHIA 8 and EPOS 3 in the 0-10$\%$ event class are given in  Fig.2.

\begin{figure}[htb!]
\begin{center}
\includegraphics[scale=0.33]{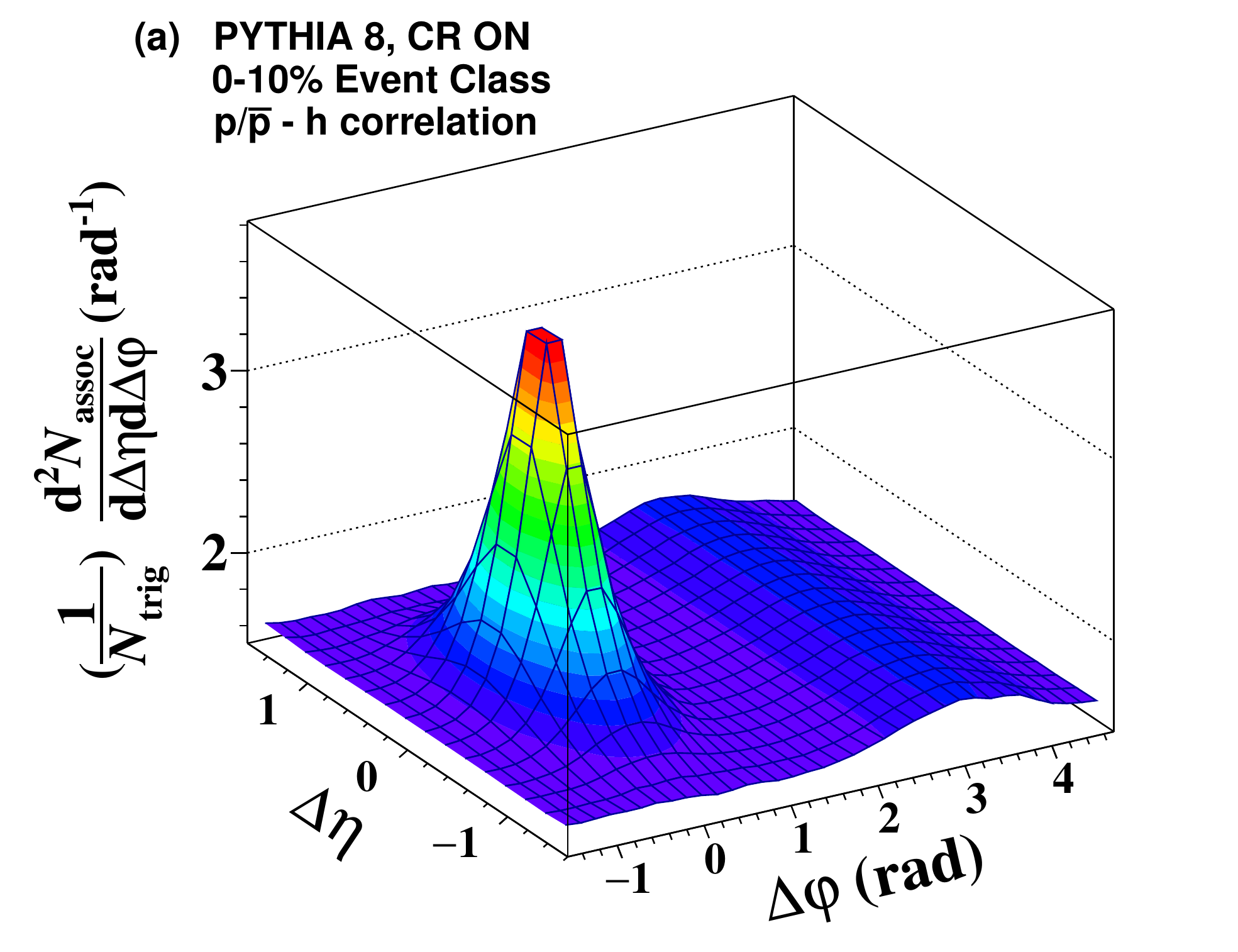}
\includegraphics[scale=0.32]{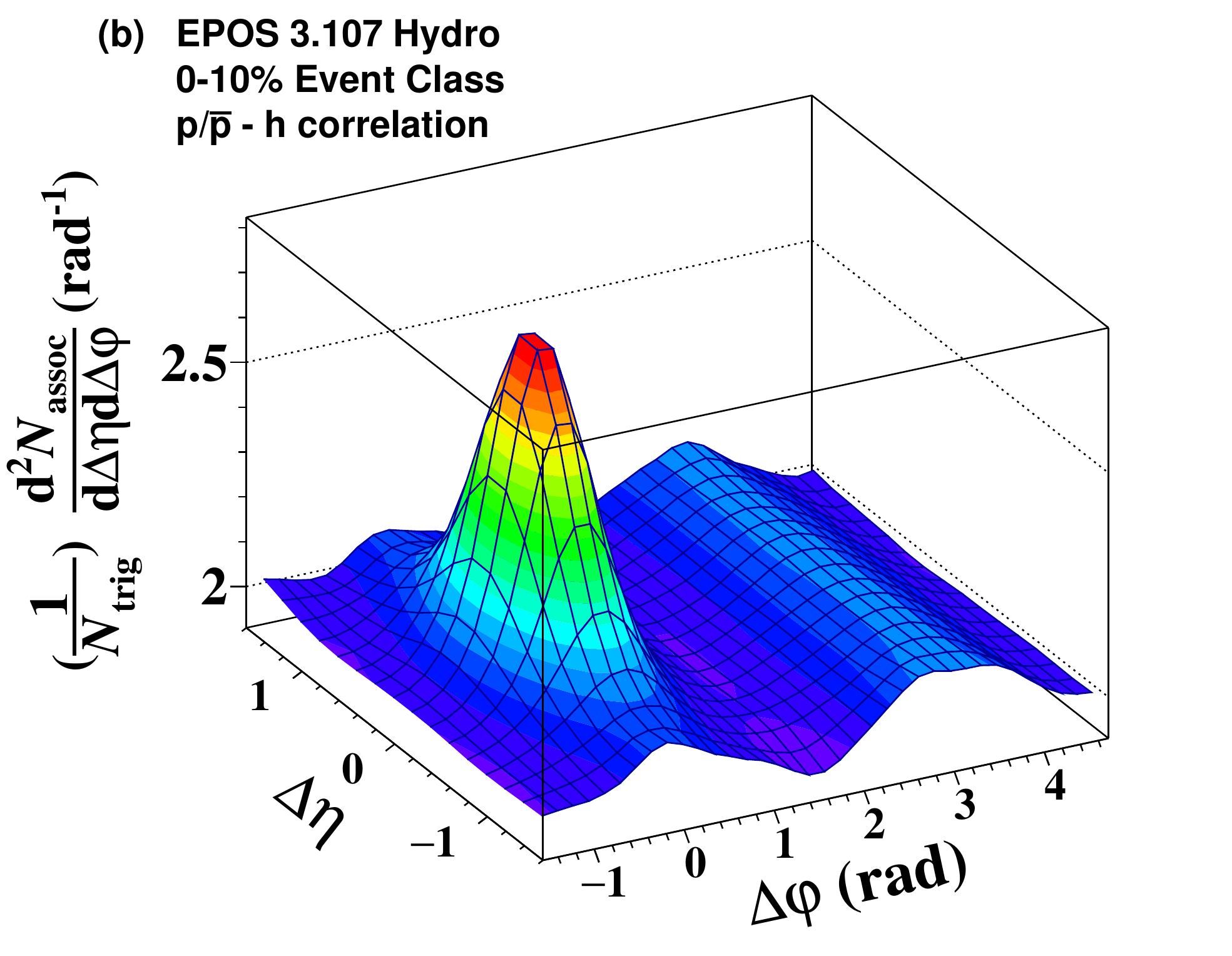}
\caption{[Color online] Two particle $\Delta\eta$-$\Delta\phi$ correlation function in 0-10 $\%$ event class 
of pp collisions at $\sqrt{s}$ = 7 TeV with proton as trigger particles from PYTHIA 8 (left) and EPOS 3 (right).}
\end{center}
\label{corr-p-pi}
\end{figure}

This analysis concentrates only on the near side ($|\Delta\phi|$  $< \pi/2$) of the correlation function.
The particles from jet fragmentation are expected to be confined in a small angular region. The flow modulated background is estimated from large $|\Delta\eta|$ ($|\Delta\eta|$ $\geq$ 1.1) and subtracted from the near side jet peak ($|\Delta\eta| <$ 1.1) as it is done in \cite {trigdilution_EPOS_pPb} \cite {MPI_pPb}. 

The $\Delta\phi$ projected correlation functions for regions $|\Delta\eta| <$ 1.1 (jet) and $ |\Delta\eta| > $1.1 (bulk) are shown in Fig 3. 
The background subtracted $\Delta\phi$ projected correlation function is shown in Fig 4. After bulk subtraction the event averaged near-side jetlike per trigger yield is calculated by integrating the
$\Delta\phi$ projection in the range  $|\Delta\phi|$  $< \pi/2$.

\begin{figure}
\begin{center}
\includegraphics[height=4.0 cm, width=6.0 cm]{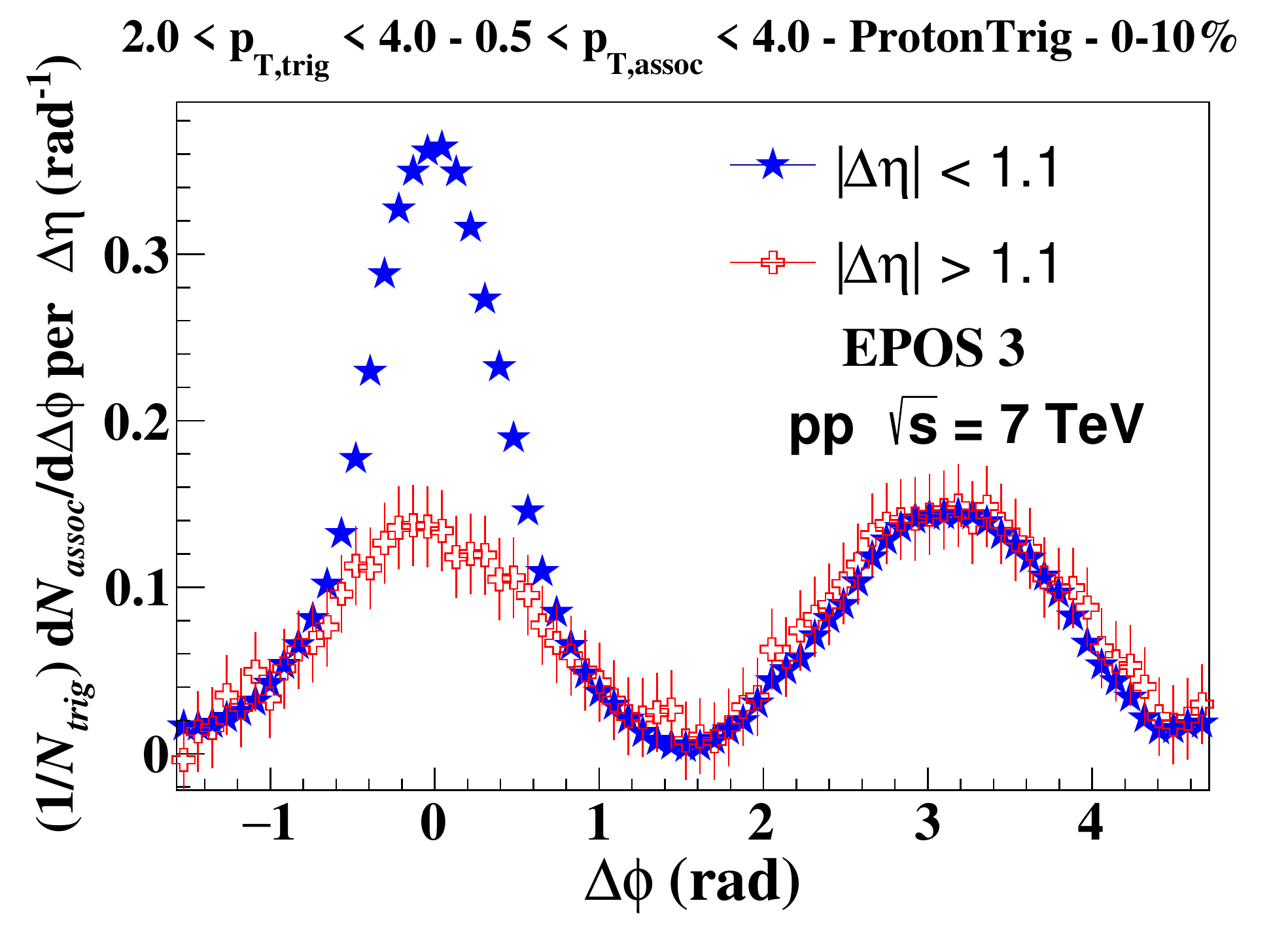}
\caption{[Color online] The  $\Delta\phi$ projected correlation function for two $\Delta\eta$ regions referred
to as jet (blue) and bulk (red).}
\label{proj-two-deleta}
\end{center}
\end{figure}

\begin{figure}
\begin{center}
\includegraphics[height=4.0 cm, width=6.0 cm]{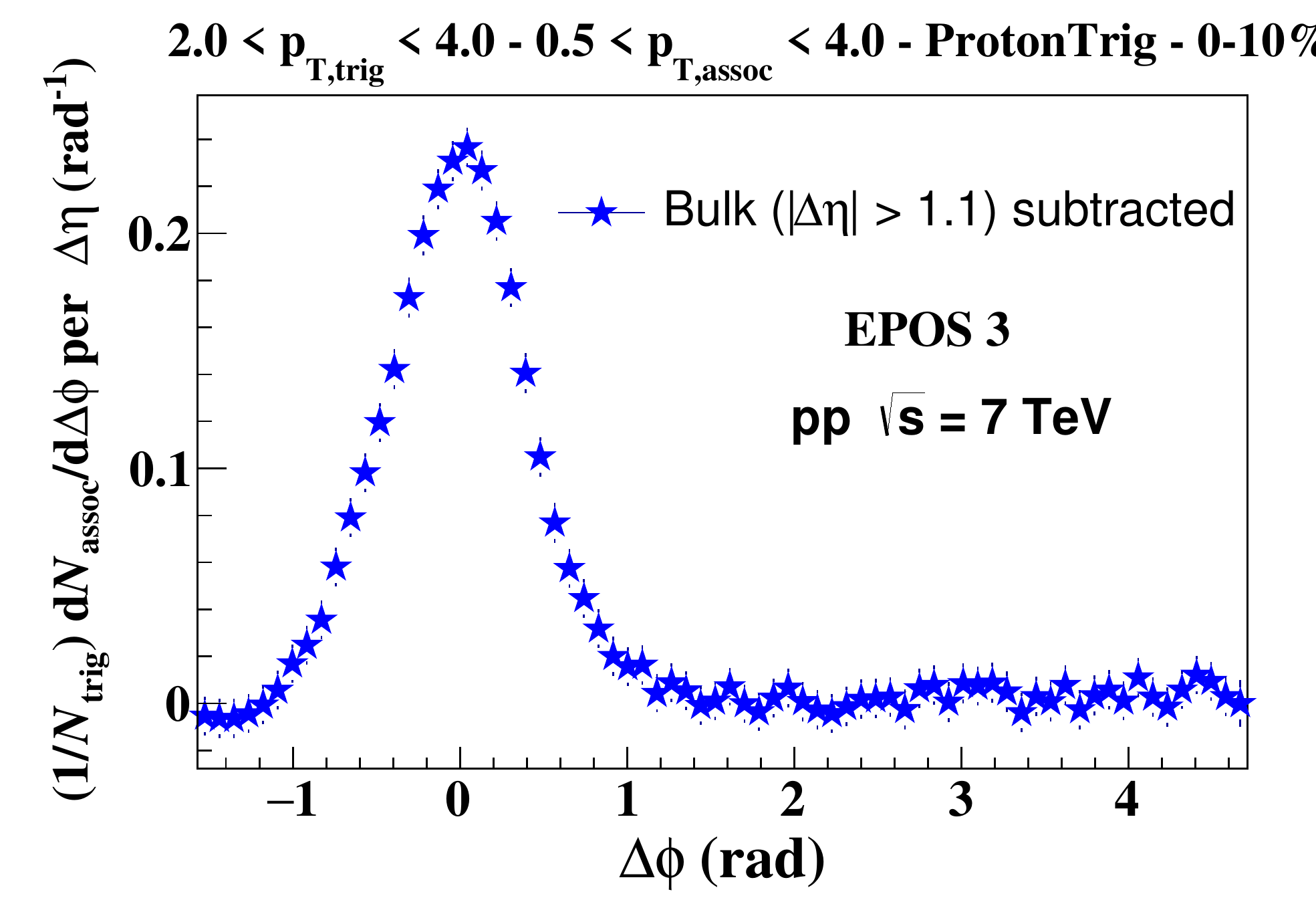}
\caption{[Color online] The  $\Delta\phi$ projected correlation function after bulk subtraction as discussed in the text.}
\label{proj-bkg-subtracted}
\end{center}
\end{figure}

\section{Results and Discussion} 
Aforementioned, in this work we measure the multiplicity evolution of the pion and proton triggered jet-like yields in pp collisions at $\sqrt{s} = $ 7 TeV. The trigger particles are selected from the intermediate $p_{T}$ range (2.0  $<p{_T}<$ 4.0 GeV/$\it{c}$) where the inclusive proton to pion enhancement has been observed \cite {alice_protontopion_SQM2016} and can be qualitatively explained by both hydrodynamics (EPOS 3) and MPI based color reconnection (PYTHIA 8) as shown in Fig 1. 

\begin{figure}[htb!]
\begin{center}
\includegraphics[scale=0.29]{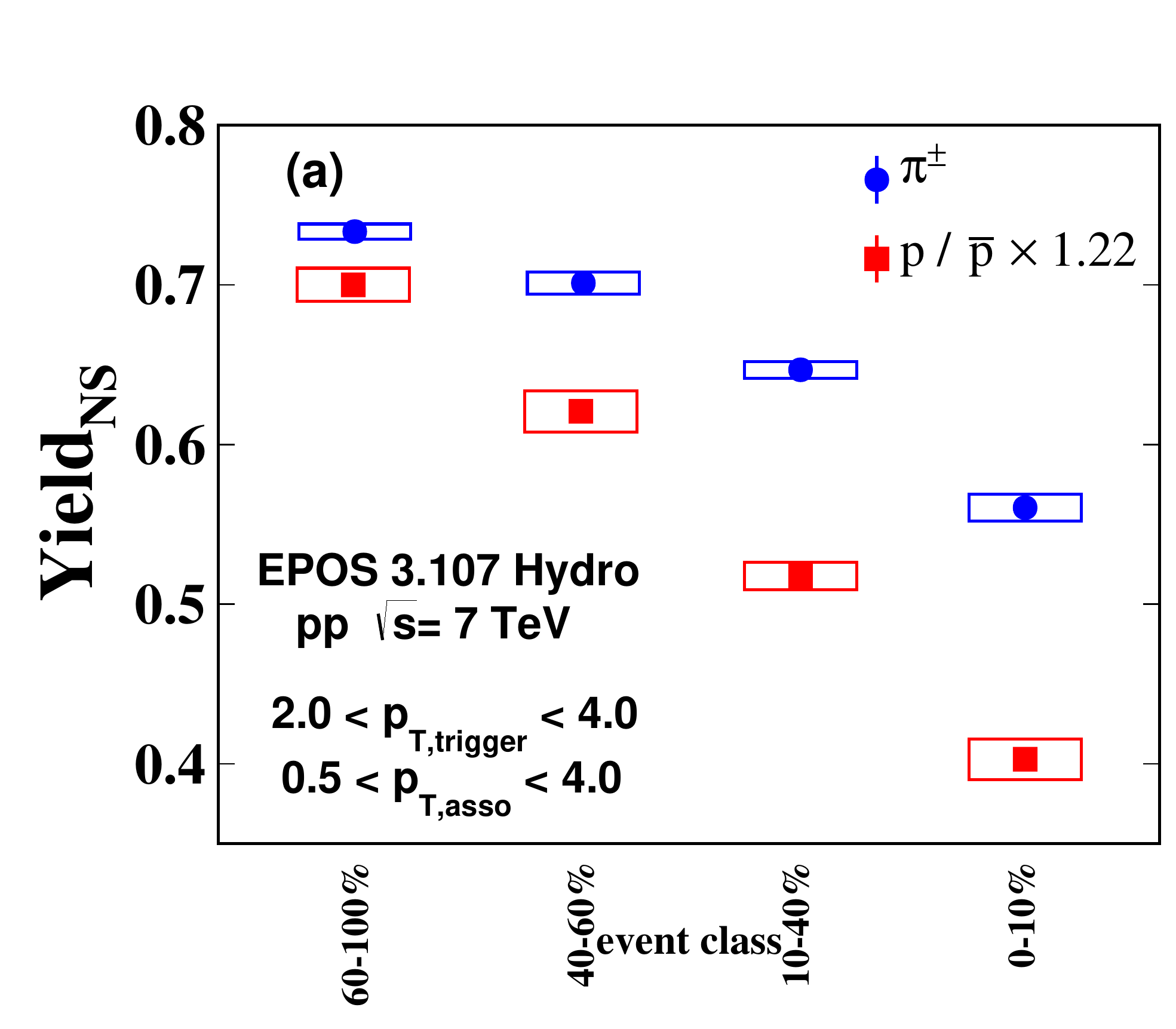}
\includegraphics[scale=0.29]{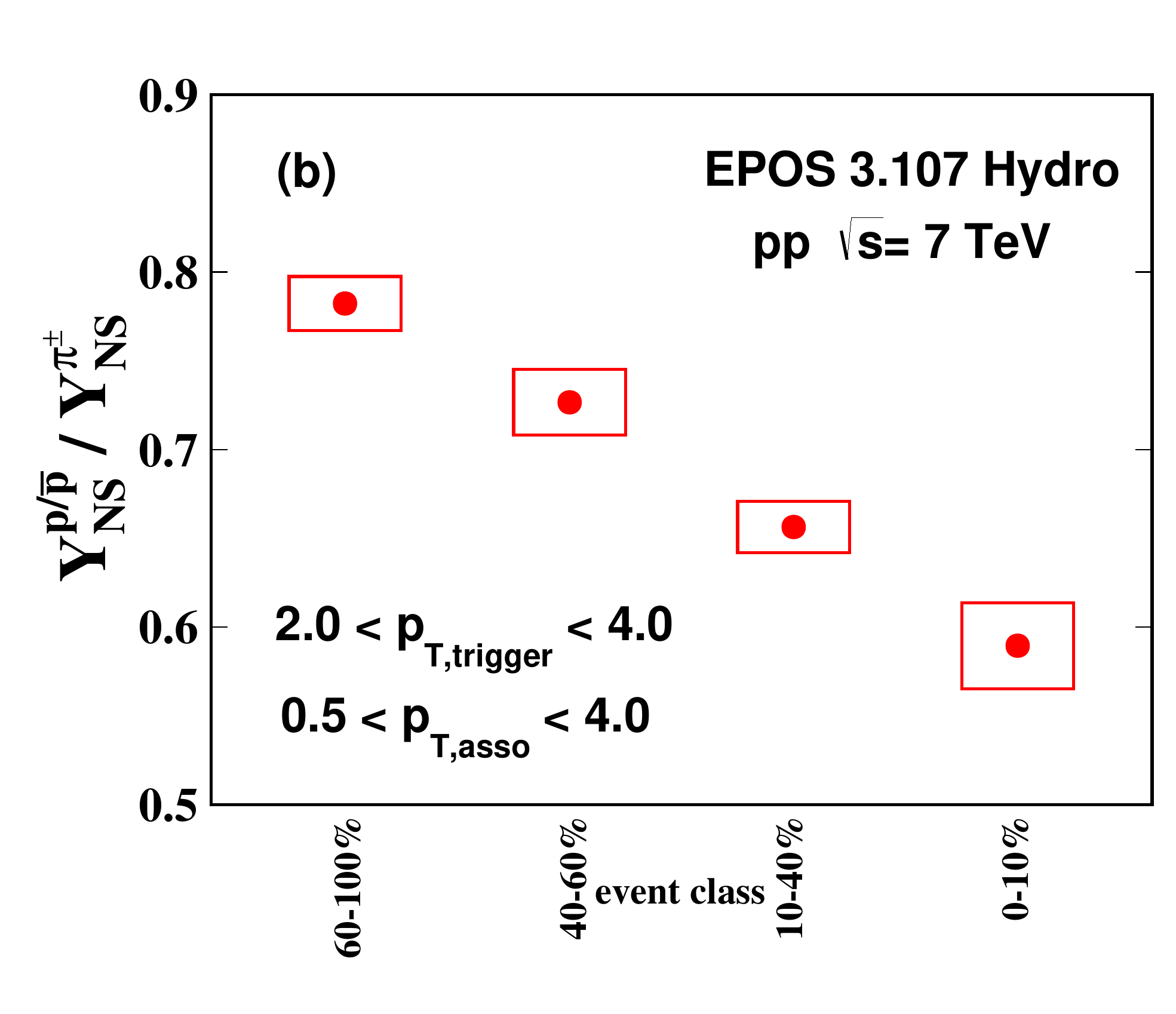}
\caption{[Color online]a) Multiplicity dependence of the near-side jet-like per trigger yield (bulk subtracted) associated with proton and pion triggers 
in pp collisions at $\sqrt{s}$ = 7 TeV from EPOS 3.107. b) Multiplicity dependence of the ratio of the proton to pion triggered yield in pp collisions at $\sqrt{s}$ = 7 TeV from EPOS 3.107.}
\end{center}
\label{Yield_ratio_Pr_Pi}
\end{figure}

\begin{figure*}[htb!]
\begin{center}
\includegraphics[scale=0.29]{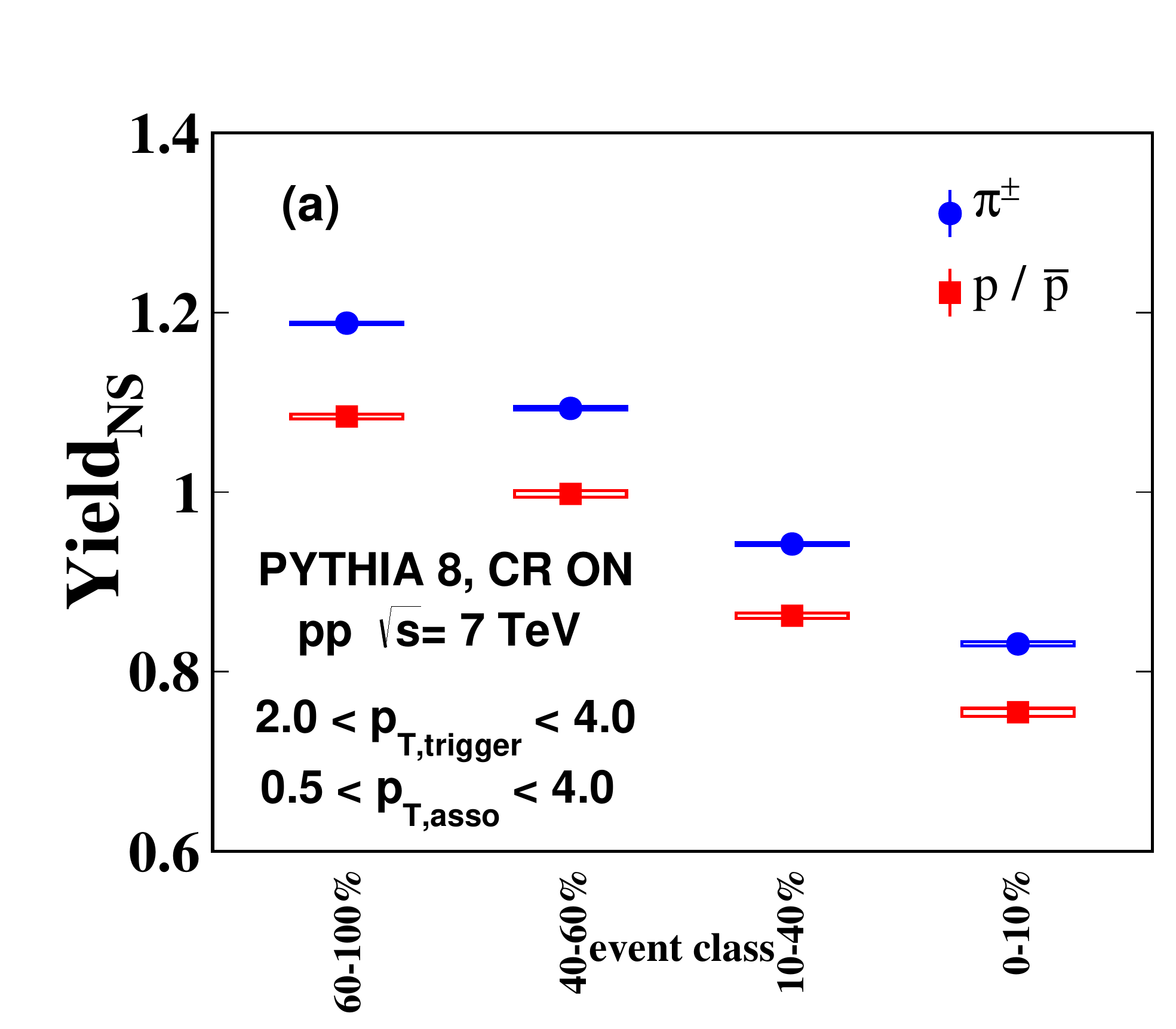}
\includegraphics[scale=0.29]{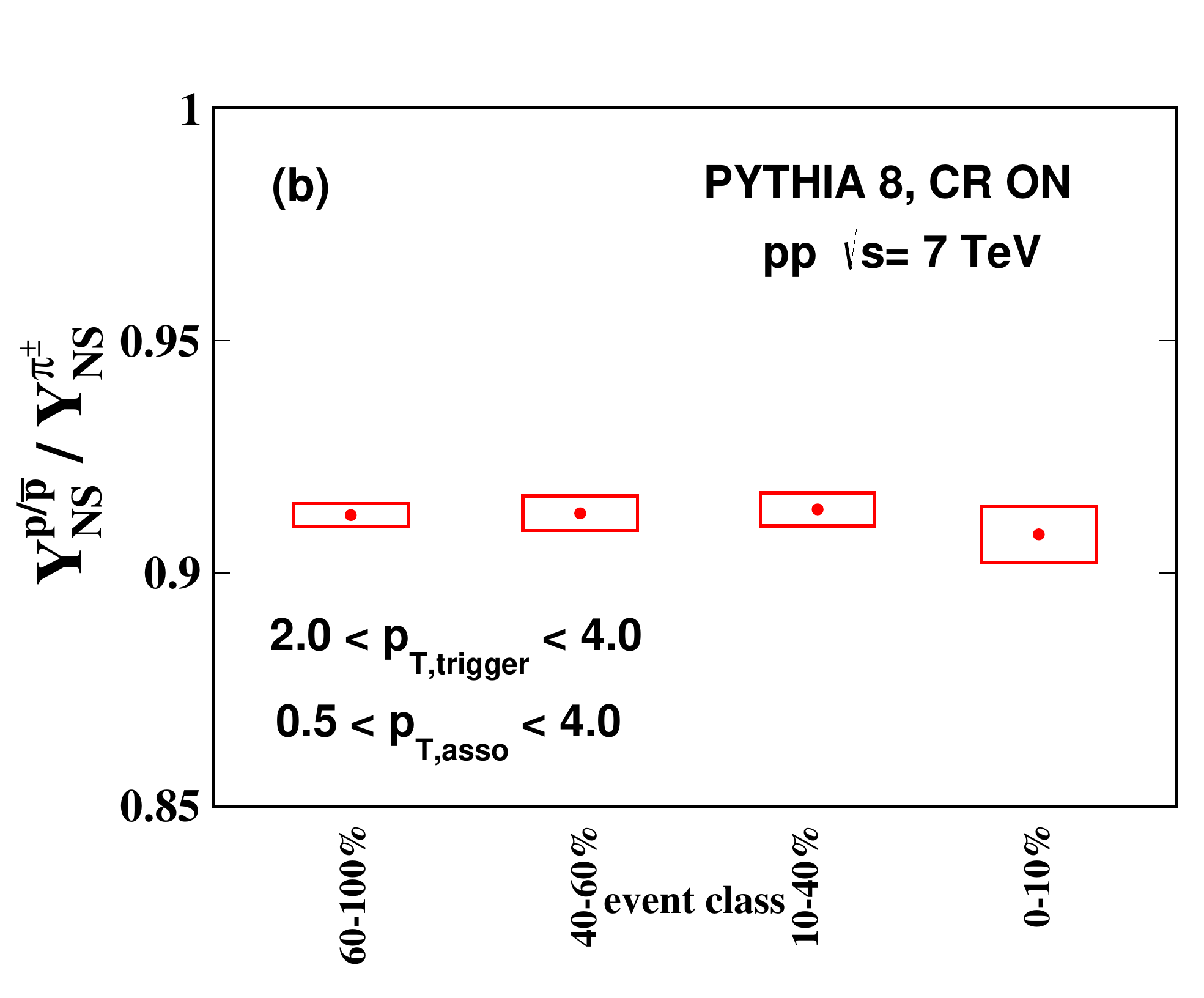}
\caption{[Color online]a) Multiplicity dependence of the near-side jet-like per trigger yield (bulk subtracted) associated with proton and pion triggers 
in pp collisions at $\sqrt{s}$ = 7 TeV from PYTHIA 8 (CR on). b) Multiplicity dependence of the ratio of the proton to pion triggered yield in pp collisions at $\sqrt{s}$ = 7 TeV from PYTHIA 8 (CR on).}
\end{center}
\label{individual_pr_pi_yield}
\end{figure*}

The multiplicity dependence of the pion and proton triggered jet-like yields as estimated from EPOS 3 is shown in Fig 5(a). The proton triggered jetlike yields decrease with multiplicity at a larger rate compared to that of pions - creating the trigger dilution \cite {trigdilution_EPOS_pPb} as shown in Fig 5(b).
In the context of EPOS model, particles originating from corona constitute the jet part of the system and create the near side jet peak around $|\Delta\eta|$ $\approx$ 0, $|\Delta\phi|$ $\approx$ 0. Whereas, the particles originating  from hydrodynamical evolution are responsible for long range ridge like structure \cite {trigdilution_EPOS_pPb} as shown in Fig 2(b). Hadrons pushed from lower to higher $p_{T}$ by radial flow  are expected not to exhibit a short range jet-like correlation beyond the expected flow (ridge) like correlation. As a result, the bulk subtracted near-side jet peak (Fig 4) is dominated by the hard triggered (jet) correlation only. Since the correlation functions are normalized by both hard and soft triggers, the soft triggers without any correlated partners in the bulk subtracted jet peak are expected to cause a dilution in the per trigger jet-like yield \cite {trigdilution_EPOS_pPb}. The rate of dilution is associated with the rate of increase in the  proportion of soft triggers with multiplicity and this rate has a trigger species dependence. As radial flow pushes massive protons more compared to lighter pions from lower to intermediate $p_{T}$ (trigger $p_{T}$ region), a stronger suppression in the baryon triggered correlation with multiplicity has been observed compared to that of mesons as shown in Fig 5 \cite{trigdilution_EPOS_pPb,phenix_triggerdilution,star_triggerdilution}. The ALICE collaboration has demonstrated a similar trigger species dependence of the dilution of jetlike yields with multiplicity in p-Pb collisions at $\sqrt{s_{NN}}$ = 5.02 TeV \cite{dsarkar_QM17}. EPOS 3 (3+1D event-by-event hydro model) and AMPT with string melting (incorporates coalescence model of hadronization) can not reproduce the data quantitatively but EPOS 3 can qualitatively mimic the multiplicity evolution of the individual pion and proton-triggered jet-like yields - indicating radial flow as a possible source of trigger dilution in p-Pb collisions at $\sqrt{s_{NN}}$ = 5.02 TeV \cite{dsarkar_QM17}.

Now, color reconnection in PYTHA 8 \cite {epos_radialflow_spectra_pPb} \cite {ortiz_radialflow} can also qualitatively explain the anomalous baryon enhancement at intermediate $p_{T}$ along with other radial flow like behaviors as observed in pp collisions at 7 TeV \cite {alice_protontopion_SQM2016}. The multiplicity evolution of the pion and proton triggered near side jet-like yields as obtained from PYTHIA 8 is shown in Fig 6(a). Both pion and proton triggered jetlike yields decrease with multiplicity but the rate of decrease has no trigger species dependence.  As PYTHIA 8 (with CR) doesn't exhibit any significant ridge like structure in high multiplicity pp collisions as shown in Fig 2(a), we argue that the decrease in the per trigger jetlike yield with multiplicity is not unique to the hydrodynamic approach or  presence of ridge like structure in the system. However, the trigger species dependence of the rate of dilution is sensitive to the underlying dynamics responsible for the baryon to meson  enhancement at the inclusive level. In EPOS 3, the baryon enhancement is due to radial flow and it generates trigger dilution effect as shown in Fig 5. Whereas, the color reconnection mechanism in PYTHIA 8 generates the anomalous bayon enhancement at the inclusive level dut doesn't exhibit any trigger species dependence of the dilution as shown in Fig 6. In absence of jet-medium interaction, the different  trend in the multiplicity evolution of pion and proton triggered jet-like correlation in EPOS 3 and PYTHIA 8 (with CR) can be used to disentangle the different underlying mechanisms (e.g hydrodynamics and color reconnection) capable of generating flow like effects in high multiplicity pp collisions at the LHC energy.

This work shows that multiplicity evolution of the identified triggered correlation in small collision systems may serve as a useful probe to investigate the origin of radial flow like effects at intermediate $p_{T}$ region of small collision systems. This observable can be used more effectively in small systems  compared to the heavy ion case where severe jet quenching affects the correlation pattern - making it difficult to disentangle the effect of soft physics (radial flow, color reconnection) from jet-medium interplay. Further comparison with the data analyzed in the way proposed in this paper will be helpful to constrain different models aiming to explain the collectve like behaviors in high multiplicity pp collisions at the LHC energies.

\ack
DS would like to acknowledge the financial support from the CBM-MUCH project grant of BI-IFCC/2016/1082(A). DS would like to thank Dr. Federico Ronchetti,  Dr. Alessandra Fantoni and Dr. Valeria Muccifora for their help and support throughout this
work. Thanks to Dr. Klaus Werner for allowing us to use EPOS 3 for this work. Thanks to VECC grid computing team for their 
constant effort to keep the facility running and helping in EPOS and PYTHIA  data generation.

\bibliography{mybibfile}

\end{document}